\documentclass[twocolumn,preprintnumbers,amsmath,amssymb,prl]{revtex4}
\usepackage{pifont}
\usepackage{amsmath}
\usepackage{amsfonts}
\usepackage{graphicx}
\usepackage{amssymb}
\usepackage{dcolumn}
\usepackage{bm}
\usepackage{slashed}

\begin{document}

\title{Modified light cone condition via vacuum polarization in a time dependent field}

\author{Huayu Hu$^{\ast}$}
\author{Jie Huang}

\affiliation{Hypervelocity Aerodynamics Institute, China Aerodynamics Research and Development Center, 621000 Mianyang, Sichuan, China}

\date{\today}

\begin{abstract}
The appearance of unconventional vacuum properties in intense fields has long been an active field of research. In this paper the vacuum polarization effect is investigated via a pump probe scheme of a probe light propagating in the vacuum excited by two counter-propagating laser beams. The modified light cone condition of the probe light is derived analytically for the situation that it passes through the electric/magnetic antinode plane of the pump field. The derivation does not follow the commonly adopted assumption of treating the pump field as a constant field. Differences from the conventional light cone conditions are identified. The implications of the result are discussed with a consideration of the vacuum birefringence measurement.
\end{abstract}

\pacs{
}

\maketitle

\subsection{Introduction}
The enthusiasm to investigate nonlinear QED properties of vacuum is significantly stimulated by the strong laser technology in progress.
Optical light with an intensity up to $2\times10^{22}$W/cm$^2$ can be generated \cite{intensity1} and the goal of an intensity beyond $10^{25}$W/cm$^2$ is set for ELI project and others \cite{intensity2}. Besides, coherent x-ray radiation of unprecedented brilliance with frequency up to $\sim10$keV is expected to be produced by free electron lasers \cite{XFEL}.
Although the proposed light intensity or photon frequency is still much smaller than the critical intensity $I_{cr}\sim10^{29}$W/cm$^2$ or Compton frequency $\omega_{c}\sim1$MeV which means vacuum cascade via spontaneous electron positron pair production would take place in such a field, it is envisaged that combined with delicate experimental design and high sensitivity measurement techniques, the direct observation of novel properties of laser-excited vacuum can be realized \cite{Homma2011, laservb, Benking, Markuland_photonacceleration}.

In quantum electrodynamics, it is predicted that electron positron pairs are spontaneously created and immediately annihilated in the vacuum due to uncertainty principle.
This has several consequences. For example, over a short time a photon can resolve into a loop of virtual electron positron pair which can couple with other photons, and thus it results in photon-photon scattering. The presence of the charges, although the lifetime of which is very short individually, collectively provides a basis for the vacuum to be polarized as a dielectric matter. Therefore it is natural to conceive pump probe technique to study the vacuum \cite{Homma2011} as that has been widely used in exploring matter physics in strong field.

The theoretical investigation to vacuum polarization is based on the effective action theory, where the high-energy degrees of freedom (the electron mass $m\sim 1$MeV) of the exact theory is integrated out and one arrives at an effective description of the low-energy degrees of freedom (the photons) which are relevant to the physics of the vacuum \cite{Dittrih_vp}.
This results in modifications to the conventional electromagnetic field lagrangian $\mathcal{L}_0=-1/4 F_{\mu\nu} F^{\mu\nu}$ with the antisymmetric field tensor $F_{\mu\nu}=\partial_\mu A_\nu-\partial_\nu A_\mu$. For laser intensity $\ll I_{cr}$ and frequency $\ll \omega_{c}$ as the case at present and in the foreseeable future, it is enough to take into account one loop correction and adopt the lowest-order Heisenberg-Euler Lagrangian $\mathcal{L}=\mathcal{L}_0+\delta\mathcal{L}$ as the effective lagrangian \cite{HE, Weisskopf, Schwinger}, with
\begin{equation}\label{HELag}
\delta\mathcal{L}=\frac{\alpha^2}{360\pi^2m^4}[\frac{1}{4}(F_{\mu\nu}F^{\mu\nu})^2+\frac{7}{16}(F_{\mu\nu}\widehat{F}^{\mu\nu})^2],
\end{equation}
where $\alpha\sim 1/137$ is the fine structure constant, and $\widehat{F}^{\mu\nu}=\epsilon^{\mu\nu\eta\tau}F_{\eta\tau}/2$ with the Levi-Civita symbol $\epsilon^{\mu\nu\eta\tau}$. Then it is common to study the vacuum polarization by assuming an electromagnetic plane wave $f^{\mu\nu}=k^\mu a^\nu-k^\nu a^\mu$ with four momentum $k^\mu$ and four potential $a^\mu$ propagating in an intense background field $\phi^{\mu\nu}$ to see how the light cone condition of $f^{\mu\nu}$ is modified \cite{Dittrih_vp, Markuland_photonacceleration}. The plane wave $f^{\mu\nu}$ serves as the probe and $\phi^{\mu\nu}$ serves as the pump field. For $\phi^{\mu\nu}$ being a constant field, mathematically rigorous results exist, that for the two polarization states $a^\mu_1\sim\widehat{\phi}^{\mu\lambda}k_\lambda$ with $\widehat{\phi}^{\mu\nu}$ being the dual tensor of $\phi^{\mu\nu}$ and $a^\mu_2\sim\phi^{\mu\lambda}k_\lambda$ the new light cone conditions are \cite{Dittrih_vp, BBfirst}
\begin{align}\label{lightcone1}
&a^\mu_1: k^2=2c_2z_k\,,\\
&a^\mu_2: k^2=2c_1z_k\,,\label{lightcone2}
\end{align}
where $c_1=2\alpha/(45\pi E_{cr}^2)$, $c_2=7\alpha/(90\pi E_{cr}^2)$, and $z_k=(k_\mu\phi^{\mu\eta})(k_\nu{\phi}^{\nu}_{\eta})$.
$E_{cr}=m^2/e=1.3\times10^{16}$V/cm is the critical electric field strength.


However, for a time dependent $\phi^{\mu\nu}$ field, explicit relations like the above which are applicable for general cases are not acquired.
Many research works have conceived a strong laser field as the background field or pump field, but in nature laser fields vary both in time and space.
A qualitative assumption is usually employed, that in studying vacuum polarization problems a field can be seen as constant as long as it varies slowly with respect to the Compton frequency, and thus for almost all relevant pump fields Eqs. (\ref{lightcone1}, \ref{lightcone2}) can be used as the light cone conditions of the probe wave \cite{Homma2011, Dittrih_vp, Markuland_photonacceleration, laservb}. In the derivation, it means that at first the total field is defined as $A^{\mu\nu}=\phi^{\mu\nu}+f^{\mu\nu}$ but later the temporal and spatial derivatives of $\phi^{\mu\nu}$ are dropped while those of $f^{\mu\nu}$ are kept. The validity of this treatment is not self-evident from a mathematical point of view. Moreover, it is long hoped that research on vacuum polarization would lead to detection of additional and unexpected signals which may herald new physics, for example the existence of lightest quark-antiquark loop, the yet undiscovered axion-like particles, and so on. Since these signals, if indeed exist, should be quite weak, it is worthwhile to understand to what extent the application of the conventional assumption would influence the accuracy of theoretical predictions in a temporally and/or spatially dependent field. In this paper, we access the problem theoretically via a pump probe scheme, where the probe light propagates in an effectively time dependent pump field. New light cone conditions are acquired analytically. They can be thought of as a direct generalization of Eqs. (\ref{lightcone1}, \ref{lightcone2}) plus an additional correction. Through the investigation of vacuum birefringence measurements, the conventional assumption is partly justified in the sense that for most cases the generalization of Eqs. (\ref{lightcone1}, \ref{lightcone2}) lead to the major contributor of the ellipticity parameter. The influence of the additional correction on the measurement is also discussed.

The system of units $\hbar=c=4\pi\epsilon_0=1$ and the metric $g=$diag$(-,+,+,+)$ are adopted in the paper.

\subsection{Light cone condition in a time dependent pump field}
Here we consider a plane wave probe light as $f^{\mu\nu}=k^\mu a^\nu-k^\nu a^\mu$ with temporal-spatial dependency $a^\mu\propto\xi(-\omega t+\vec{k}\cdot\vec{x})$ propagating in a vacuum excited by a time dependent pump field $\phi^{\mu\nu}$. A time varying but space homogeneous field can be realized in the vicinity of an antinode plane of a standing wave formed by two counterpropagating electromagnetic waves. Assume for simplicity the waves are linearly polarized in $\hat{e}_x$ direction, the wave vectors are in the directions of  $\pm\hat{e}_z$, and the electric components are respectively $\mathcal{E}/2 \hat{e}_x\cos(\omega_\phi t-k_\phi z)$ and $\mathcal{E}/2 \hat{e}_x\cos(\omega_\phi t+k_\phi z)$. Then the total electric component of the pump field is $\vec{E}=\mathcal{E}\hat{e}_x\cos k_\phi z \cos\omega_\phi t$ and the total magnetic component is $\vec{B}=\mathcal{E}\hat{e}_y\sin k_\phi z\sin\omega_\phi t$. The electric antinode planes are those with fixed $z$ value to satisfy $\sin k_\phi z=0$. Along the path of the probe beam on such antinode plane there is a spatially uniform electric field $\vec{E}=\mathcal{E}\hat{e}_x \cos\omega_\phi t$.  Such time dependent electric field scenario is often adopted in studying strong field pair production problems \cite{antinode, electricfield, electricfield2}. Note that although the amplitude of the magnetic field vanishes on the plane, its spatial derivative does not.

To be consistent with the conditions of the effective action approach, $\omega_c\gg\omega\gg\omega_\phi$ is required. The second half of the inequality comes from the consideration that the probe beam waist size which is larger than its wavelength should be much smaller than the wavelength of the pump field, in order to assure that the crosssection of the probe beam locates in the close vicinity of the antinode. The variation of the effective lagrangian with respect to the four-potential $A^\mu$ gives
\begin{equation}\label{HEWav}
0=\partial_\mu[F^{\mu\nu}-\frac{c_1}{2}(F_{\mu\nu}F^{\mu\nu})F^{\mu\nu}-\frac{c_2}{2}(F_{\mu\nu}\widehat{F}^{\mu\nu})\widehat{F}^{\mu\nu}]\,,
\end{equation}
where the total field strength tensor is $F^{\mu\nu}=\phi^{\mu\nu}+f^{\mu\nu}$. As illustrated in previous studies \cite{Dittrih_vp}, the self interaction of the plane wave can be neglected, regarding the plane wave as reminiscent of a test charge in classical electrodynamics. Besides, considering the probe light intensity to be much weaker than that of the pump field, the wave equation can be linearized with respect to $f^{\mu\nu}$. In the following derivation we would not employ the constant field assumption but keep the derivatives of $\phi^{\mu\nu}$, to find out explicitly what modifications to the light cone condition would turn out to be.

The linearization of the wave equation results in
\begin{align} \label{linear_wave}
0=\partial_\mu f^{\mu\nu}&-\frac{c_1}{2}\partial_\mu(\phi^2 f^{\mu\nu}+2\phi_{\alpha\beta}f^{\alpha\beta}\phi^{\mu\nu})\nonumber\\
&-\frac{c_2}{2}\partial_\mu(\phi_{\alpha\beta}\widehat{\phi}^{\alpha\beta}\widehat{f}^{\mu\nu}+2\widehat{\phi}_{\alpha\beta}f^{\alpha\beta}\widehat{\phi}^{\mu\nu}) \,.
\end{align}
Making use of the Bianchi identity $\partial_\mu\widehat{f}^{\mu\nu}=0$, $\partial_\mu\widehat{\phi}^{\mu\nu}=0$,  and noticing that under the specified configuration of $\phi^{\mu\nu}$ there is  $\mathcal{G}=\frac{1}{4}\phi_{\mu\nu}\widehat{\phi}^{\mu\nu}=-\vec{E}\cdot\vec{B}=0$, the equation can be written as
\begin{align} \label{linear_wave2}
0&=\{(1-2c_1\mathcal{F})\partial_\mu f^{\mu\nu}-c_1\phi_{\alpha\beta}\phi^{\mu\nu}\partial_\mu f^{\alpha\beta}-c_2\widehat{\phi}_{\alpha\beta}\widehat{\phi}^{\mu\nu}\partial_\mu f^{\alpha\beta}\}\nonumber\\
&+\{-2 c_1f^{\mu\nu}\partial_\mu \mathcal{F} -c_1 f^{\alpha\beta}\partial_\mu(\phi_{\alpha\beta}\phi^{\mu\nu})-c_2f^{\alpha\beta}\hat{\phi}^{\mu\nu}\partial_\mu\hat{\phi}_{\alpha\beta}\}\,,
\end{align}
where $\mathcal{F}=\frac{1}{4}\phi_{\mu\nu}\phi^{\mu\nu}$. The expression in the first brace is exempted from derivatives of the pump field $\phi^{\mu\nu}$ and leads to the light cone conditions similar to (\ref{lightcone1}, \ref{lightcone2}). The expression in the second brace leads to additional corrections.

Performing calculations on the electric antinode plane yields
\begin{align}
&\mathcal{F}=-\frac{1}{2}\mathcal{E}^2\cos^2{\omega_\phi t},\nonumber\\
&\partial_i \mathcal{F}=0\; {\tt for} \; i\neq 0,\nonumber\\
&\phi^{\mu\nu}\partial_\mu \phi_{\alpha\beta}=\phi^{0\nu}\partial_0 \phi_{\alpha\beta},\nonumber\\
&\hat{\phi}^{\mu\nu}\partial_\mu\hat{\phi}_{\alpha\beta}=\hat{\phi}^{3\nu}\partial_3\hat{\phi}_{\alpha\beta},\nonumber\\
&\partial_\mu\phi^{\mu\nu}=\partial_0\phi^{0\nu}+\partial_3\phi^{3\nu}.
\end{align}
Define three antisymmetric constant tensors $h^{\mu\nu}$, $\phi_E^{\mu\nu}$ and $\phi_B^{\mu\nu}$ as $f^{\mu\nu}=h^{\mu\nu}\xi(-\omega t+\vec{k}\cdot\vec{x})$, $\phi^{0\nu}=\phi_E^{0\nu}\cos k_\phi z \cos\omega_\phi t$ with $\phi_E^{ij}=0$ for $i,j=1,2,3$, and $\phi^{ij}=\phi_B^{ij}\sin k_\phi z\sin\omega_\phi t$ with $i,j=1,2,3$ while $\phi_B^{0\nu}=0$.
Similarly a constant four-vector $a^\mu_h$ can be defined as $a^\mu=a^\mu_h\xi(-\omega t+\vec{k}\cdot\vec{x})$, so that $h^{\mu\nu}=k^\mu a_h^\nu-k^\nu a_h^\mu$.
Besides, denote $\mathcal{G}_E=\frac{1}{4}\phi_{E,\mu\nu}\widehat{\phi}_E^{\mu\nu}=0$ and $\mathcal{F}_E=\frac{1}{4}\phi_{E,\mu\nu}\phi_E^{\mu\nu}=-\frac{1}{2}\mathcal{E}^2$.
We can get the equation
\begin{align} \label{linear_wave3}
0&=\{(1-2b_1\mathcal{F}_E)k_\mu h^{\mu\nu}-b_1\phi_{E,\alpha\beta}\phi_E^{\mu\nu}k_\mu h^{\alpha\beta}\nonumber\\
&-b_2\widehat{\phi}_{E,\alpha\beta}\widehat{\phi}_E^{\mu\nu}k_\mu h^{\alpha\beta}\}
+\Gamma_1 \Gamma_2 \omega_\phi \{- 4 b_1 \mathcal{F}_E h^{0\nu}\nonumber\\
&-b_1 h^{\alpha\beta}\phi_E^{0\nu}\phi_{E,\alpha\beta}-b_1 h^{\alpha\beta}\phi_{E,\alpha\beta}(\phi_E^{0\nu}-\phi_B^{3\nu})\nonumber\\
&+ b_2 h^{\alpha\beta}\hat{\phi}_E^{3\nu} \hat{\phi}_{B,\alpha\beta}\}\,,
\end{align}
where $\Gamma_1=\xi/\xi'$ with $\xi'$ the first order derivative of $\xi$, $\Gamma_2=-\sin\omega_\phi t/\cos{\omega_\phi t}$, and $b_{1,2}=c_{1,2}\cos^2{\omega_\phi t}$.

Try the polarization state $a_h^\mu\sim\widehat{\phi}_E^{\mu\lambda}k_\lambda=\mathcal{E}\hat{e}_x\times\vec{k}\sim\hat{e}_z$. According to the configuration of the setup, several identities exist such as $\widehat{\phi}_E^{0\lambda}=0$, $\phi_{E,\mu\eta}\widehat{\phi}_E^{\eta\nu}=-\mathcal{G}_E \delta_{\mu\nu}=0$ and $k_3=0$ since the path of the probe wave is perpendicular to the $z$ axis.  They lead to the following relations :
\begin{align}\label{identity1}
h^{\alpha\beta}\phi_{E,\alpha\beta}&=(k^\alpha \widehat{\phi}_E^{\beta\lambda}k_\lambda-k^\beta \widehat{\phi}_E^{\alpha\lambda}k_\lambda)\phi_{E,\alpha\beta}\nonumber\\
&=-2\mathcal{G}_{E} k^2\nonumber\\
&=0,\\
h^{\alpha\beta}\hat{\phi}_{B,\alpha\beta}&=(k^\alpha \widehat{\phi}_E^{\beta\lambda}k_\lambda-k^\beta \widehat{\phi}_E^{\alpha\lambda}k_\lambda)\hat{\phi}_{B,\alpha\beta}\nonumber\\
&=2(k^0 \widehat{\phi}_E^{\beta\lambda}k_\lambda-k^\beta \widehat{\phi}_E^{0\lambda}k_\lambda)\hat{\phi}_{B,0\beta}\nonumber\\
&=2k^0 \widehat{\phi}_E^{23}k_3\hat{\phi}_{B,02}\nonumber\\
&=0. \label{identity2}
\end{align}
Further making use of the identity $\phi_E^{\mu\eta}{\phi}^\nu_{E,\eta}-\widehat{\phi}_E^{\mu\eta}\widehat{\phi}^\nu_{E, \eta}=2 \mathcal{F}_E g^{\mu\nu}$ \cite{Dittrih_vp}, the equation yields
\begin{align}\label{disp_eq}
&0=\{[(1-2b_1\mathcal{F}_E+4b_2\mathcal{F}_E)k^2-2b_2 z_k](\widehat{\phi}_E^{\nu\lambda}k_\lambda)\nonumber\\
&-(2b_1\mathcal{G}_E k^2)(\phi_E^{\nu\lambda}k_\lambda)\}+\Gamma_1\Gamma_2\omega_\phi \{-4b_1 \omega\mathcal{F}_E (\widehat{\phi}_E^{\nu\lambda}k_\lambda)\}\,,
\end{align}
where $z_k=(k_\mu\phi_E^{\mu\eta})(k_\nu{\phi}^{\nu}_{E,\eta})=\omega^2\mathcal{E}^2\sin^2\theta_E$ with $\theta_E$ being the angle between the directions of the probe light momentum $\vec{k}$ and the electric pump field $\vec{E}$. It can be seen that the additional terms (the second brace) only contain couplings between the electric pump field and the electric component of the probe wave.

Since the deviation of the modified light cone condition from the trivial one is expected to be in the first order of $b_{1, 2}$, that $k^2=0+\mathcal{O}(b_1, b_2)$ \cite{Dittrih_vp}, keeping in Eq. (\ref{disp_eq}) the terms up to the first order of $b_{1, 2}$, there is
\begin{align}\label{disp_eq2}
0=&[k^2-2b_2 z_k-4b_1\Gamma_1\Gamma_2\omega_\phi \omega \mathcal{F}_E](\widehat{\phi}^{\nu\lambda}k_\lambda)\nonumber\,,\\
0=&k^2-2b_2 (1-\Gamma_1\Gamma_2\frac{4\,\omega_\phi}{7\omega\sin^2\theta_E})z_k\,.
\end{align}
As $\omega_\phi\rightarrow 0$ corresponding to a constant pump field, the documented light cone condition (\ref{lightcone1}) is recovered. But generally speaking, accounting for the time varying of the pump field results in a notable modification to it.

Then try the other polarization state $a_h^\mu\sim \phi_E^{\mu\lambda}k_\lambda$. There are
\begin{align}\label{identity3}
h^{\alpha\beta}\phi_{E,\alpha\beta}&=(k^\alpha \phi_E^{\beta\lambda}k_\lambda-k^\beta \phi_E^{\alpha\lambda}k_\lambda)\phi_{E,\alpha\beta}\nonumber\\
&=-2z_k,\\
h^{\alpha\beta}\hat{\phi}_{B,\alpha\beta}&=(k^\alpha \phi_E^{\beta\lambda}k_\lambda-k^\beta \phi_E^{\alpha\lambda}k_\lambda)\hat{\phi}_{B,\alpha\beta}\nonumber\\
&=2(k^0 \phi_E^{2\lambda}k_\lambda-k^2 \phi_E^{0\lambda}k_\lambda)\hat{\phi}_{B,02}\nonumber\\
&=-2k^2 \phi_E^{0\lambda}k_\lambda\hat{\phi}_{B,02},\label{identity4}
\end{align}
and here $k^2$ means the projection of the wave vector on the $\hat{e}_y$ direction, denoting it as $k_y$ in the following. We can get
\begin{align}\label{disp_eq3}
0=&\{[(1-2b_1\mathcal{F}_E)k^2-2 b_1 z_k](\phi_E^{\nu\lambda}k_\lambda)\nonumber\\
&-2b_2\mathcal{G}_E k^2(\widehat{\phi}_E^{\nu\lambda}k_\lambda)\}+\Gamma_1\Gamma_2\omega_\phi \{-4b_1\omega\mathcal{F}_E (\phi_E^{\nu\lambda}k_\lambda)\nonumber\\
&+4b_1\mathcal{F}_E k^\nu \phi_E^{0\lambda}k_\lambda+4b_1 z_k\phi_E^{0\nu}\nonumber\\
&-2b_1z_k\phi_B^{3\nu}+b_2(-2k_yk_\lambda\phi_E^{0\lambda}\widehat{\phi}_{B,02})\widehat{\phi}_E^{3\nu}\}\,.
\end{align}
Dot multiply both sides of the equation by $k_\nu$, and this yields $0=(8\mathcal{F}_E k^2 +8 z_k-7 k_y \widehat{\phi}_{B,02}\widehat{\phi}_E^{3\nu}k_\nu)\phi_E^{0\lambda}k_\lambda-4z_k\phi_B^{3\nu}k_\nu$. This equation is validated when $\phi_E^{0\lambda}k_\lambda=0$ and $\phi_B^{3\nu}k_\nu=0$. Both conditions are fulfilled simultaneously when $\vec{k}\cdot\vec{E}=0$ or in other words when the probe wave is propagating in $\pm\hat{e}_y$ direction. Therefore the assumption of $\phi_E^{\mu\lambda}k_\lambda$ serving as a polarization state works under the condition that $\vec{k}\cdot\vec{E}=0$. Assume this requirement is satisfied, thus $\theta_E=\pm\pi/2$. The explicit form of the polarization state is $a_h^\mu\sim \phi_E^{\mu\lambda}k_\lambda=\omega\mathcal{E}\hat{e}_x$ and there exist several identities like $\phi_E^{0\nu}=(0,\mathcal{E},0,0)=\phi_E^{\nu\lambda}k_\lambda/\omega$ and $\phi_B^{3\nu}=(0,\mathcal{E},0,0)=\phi_E^{\nu\lambda}k_\lambda/\omega$. Therefore,
\begin{align}\label{disp_eq4}
0=&\{[(1-2b_1\mathcal{F}_E)k^2-2 b_1 z_k](\phi_E^{\nu\lambda}k_\lambda)\nonumber\\
&-2b_2\mathcal{G}_E k^2(\widehat{\phi}_E^{\nu\lambda}k_\lambda)\}+\Gamma_1\Gamma_2\omega_\phi \{(-4b_1\omega\mathcal{F}_E \nonumber\\
&+4b_1 z_k/\omega-2b_1z_k/\omega)(\phi_E^{\nu\lambda}k_\lambda)\}\,.
\end{align}
Similarly to the aforementioned argument, keeping only the terms to the first order of $b_{1,2}$, Eq.\,(\ref{disp_eq4}) can be simplified as
\begin{equation}
0=[k^2-2 b_1 z_k+\Gamma_1\Gamma_2\omega_\phi (-4b_1\omega\mathcal{F}_E+2b_1 z_k/\omega)](\phi^{\nu\lambda}k_\lambda)\,,
\end{equation}
and the light cone condition is
\begin{equation}\label{disp_eq5}
0=k^2-2 b_1 (1-2\Gamma_1\Gamma_2\frac{\omega_\phi}{\omega})z_k\,.
\end{equation}
As $\omega_\phi\rightarrow 0$, light cone condition (\ref{lightcone2}) is recovered. It is interesting to note that in this polarization state the magnetic field contributes through the term $z_k\phi_B^{3\nu}$ in Eq. (\ref{disp_eq3}), although the amplitude of the magnetic field vanishes along the path of the probe light.


To go beyond the constraint $\vec{k}\cdot\vec{E}=0$,  we resort to the treatment devised in \cite{Dittrih_vp} to obtain a light cone condition averaged over the polarization states of the probe light, while the polarization states do not need to be specified out explicitly. Substituting $h^{\mu\nu}=k^\mu a_h^\nu-k^\nu a_h^\mu$ into Eq. (\ref{linear_wave3}),
multiplying both sides by ${a_h}_\nu$ and summing over the two polarization states according to the general relation $\sum_{\tt pol.} a_h^\mu {a_h}_\nu\sim g^\mu_\nu$,
this yields
\begin{equation}
0=2k^2(1+2b_2\mathcal{F}_E)-2(b_1+b_2)z_k+2\Gamma_1\Gamma_2\frac{\omega_\phi}{\omega\sin^2\theta_E}(2b_1+b_2)z_k \,.
\end{equation}
The averaged light cone condition is
\begin{align}\label{avlc}
0=k^2-(b_1+b_2-3 b_1\Gamma_1\Gamma_2 \frac{\omega_\phi}{\omega\sin^2\theta_E})z_k& \nonumber\\
+(b_2-b_1)\Gamma_1\Gamma_2 \frac{\omega_\phi}{\omega\sin^2\theta_E}z_k& \,,
\end{align}
where the first two terms on the right hand side can be recognized as plainly the arithmetic average of the light cone conditions (\ref{disp_eq2}) and (\ref{disp_eq5}). The last term appears due to the involvement of polarization states oriented in directions other than $\hat{e}_z$ and $\hat{e}_x$. Besides, the $b_2$ coefficient in the last term originates directly from the term $\sim b_2 h^{\alpha\beta} \hat{\phi}_E^{3\nu} \hat{\phi}_{B,\alpha\beta}$ in Eq. (\ref{linear_wave3}) and thus again manifests the influence of the magnetic field.

So far the probe light traveling on the electric antinode plane is considered. Derivations can be performed also for the probe light traveling on the magnetic antinode plane, where the pump field is in effect a time varying magnetic field. In order to compare with the above results, taking $k=(\omega,0,k_y,0)$, we can get an equation similar to Eq. (\ref{linear_wave3}), that
\begin{align} \label{mag1}
0&=\{(1-2b_3\mathcal{F}_B)k_\mu h^{\mu\nu}-b_3\phi_{B,\alpha\beta}\phi_B^{\mu\nu}k_\mu h^{\alpha\beta}\nonumber\\
&-b_4\widehat{\phi}_{B,\alpha\beta}\widehat{\phi}_B^{\mu\nu}k_\mu h^{\alpha\beta}\}
+\Gamma_1 \Gamma_3 \omega_\phi \{- 4 b_3 \mathcal{F}_B h^{0\nu} \nonumber\\
&+b_3 h^{\alpha\beta}\phi_B^{3\nu}\phi_{E,\alpha\beta} -b_3 h^{\alpha\beta}\phi_{B,\alpha\beta}(\phi_B^{0\nu}-\phi_E^{3\nu})\nonumber\\
&- b_4 h^{\alpha\beta}\hat{\phi}_B^{0\nu} \hat{\phi}_{B,\alpha\beta}\}\,,
\end{align}
where $\Gamma_3=\cos{\omega_\phi t}/\sin\omega_\phi t$, $\mathcal{F}_B=\frac{1}{2}\mathcal{E}^2$ and $b_{3,4}=c_{1,2}\sin^2{\omega_\phi t}$.

For the polarization state $a_h^\mu\sim\widehat{\phi}_E^{\mu\lambda}k_\lambda\sim\hat{e}_z$, besides the identities (\ref{identity1}) and (\ref{identity2}) there is also
\begin{align}
h^{\alpha\beta}\phi_{B,\alpha\beta}&=(k^\alpha \widehat{\phi}_E^{\beta\lambda}k_\lambda-k^\beta \widehat{\phi}_E^{\alpha\lambda}k_\lambda)\phi_{B,\alpha\beta}\nonumber\\
&=2(k^1 \widehat{\phi}_E^{3\lambda}k_\lambda-k^3 \widehat{\phi}_E^{1\lambda}k_\lambda)\phi_{B,13}\nonumber\\
&=0.
\end{align}
Substituting the identities into the equation, it becomes
\begin{equation}
0=(1-2b_3\mathcal{F}_B)k^2\widehat{\phi}_E^{\mu\lambda}k_\lambda-2b_3\Gamma_1\Gamma_3\frac{\omega_\phi}{\omega}z_k\widehat{\phi}_E^{\mu\lambda}k_\lambda\,.
\end{equation}
Accordingly the light cone condition is
\begin{equation}\label{malc}
0=k^2-2b_3\Gamma_1\Gamma_3\frac{\omega_\phi}{\omega}z_k\,,
\end{equation}
with $z_k=\omega^2\mathcal{E}^2$ since $\theta_E=\pi/2$.

For the polarization state $a_h^\mu\sim \phi_E^{\mu\lambda}k_\lambda\sim\hat{e}_x$, in addition to Eqs. (\ref{identity3}, \ref{identity4}), there is also
\begin{align}
h^{\alpha\beta}\phi_{B,\alpha\beta}&=(k^\alpha \phi_E^{\beta\lambda}k_\lambda-k^\beta \phi_E^{\alpha\lambda}k_\lambda)\phi_{B,\alpha\beta}\nonumber\\
&=2(k^1 \phi_E^{3\lambda}k_\lambda-k^3 \phi_E^{1\lambda}k_\lambda)\phi_{B,13}\nonumber\\
&=0.
\end{align}
Using these identities, the equation becomes
\begin{align}
0&=(1-2b_3\mathcal{F}_B)k^2 \phi_E^{\nu\lambda}k_\lambda
+\Gamma_1 \Gamma_3 \omega_\phi \{- 2 b_3 \frac{z_k}{\omega} \phi_E^{\nu\lambda}k_\lambda \nonumber\\
&-2b_3\frac{z_k}{\omega} \phi_E^{\nu\lambda}k_\lambda \}\,.
\end{align}
In accordance, the light cone condition is
\begin{equation}\label{malc2}
0=k^2-4b_3\Gamma_1\Gamma_3\frac{\omega_\phi}{\omega}z_k\,.
\end{equation}

As $\omega_\phi\rightarrow 0$, the light cone conditions (\ref{malc}, \ref{malc2}) reduce to the trivial one instead of the conditions (\ref{lightcone1}, \ref{lightcone2}). This is due to our choice of the probe light to be $\vec{k}\cdot\vec{B}=0$. If the direction of the probe light wave vector is along $\pm \hat{e}_x$,  the conditions (\ref{lightcone1}, \ref{lightcone2}) would be regained as $\omega_\phi\rightarrow 0$. It shows the strong dependency of the light cone condition on the particular situation, like the relative directions between the pump and probe field.

The analytical results (\ref{disp_eq2}, \ref{disp_eq5}, \ref{avlc}, \ref{malc} and \ref{malc2}) display explicitly that the light cone conditions of the probe light in a time varying pump field are different from that in a constant pump field. Besides the analytical results, a phenomenological explanation is attempted. The deviation of the light cone condition from the trivial one $k^2=0$ is due to the presence of virtual electron-positron pairs with a lifetime $\sim1/\omega_c$. On the one hand, one virtual pair can not feel the temporal change of the pump field with frequency $\ll\omega_c$, and should see a static $\vec{E}(t)$ or $\vec{B}(t)$ field at any given time $t$.  On the other hand, during the characteristic formation time of the probe light which scales as $\sim 1/\omega$, the occurrence time of the virtual loop is uncertain. This in principle exerts an averaged impact on the probe wave dynamics due to the time varying of the pump field.

Next, the influence of the modifications in measurements is discussed.

\subsection{Vacuum birefringence measurement}
Consider a beam of probe light propagating in the $\hat{e}_y$ direction on the electric antinode plane of a pump field $\phi^{\mu\nu}$ defined above. At first sight, it seems that the light cone conditions (\ref{disp_eq2}, \ref{disp_eq5}) indicate that different phase planes of the probe wave propagate at different velocities, since the indexes of refraction $n=|\vec{k}|/\omega$ for different polarization states are, respectively
\begin{align}\label{dispt1}
&\hat{e}_z:\;\;n_1\simeq 1+b_2 (1-\Gamma_2\frac{4\,\omega_\phi \xi(\varphi)}{7\;\omega \xi'(\varphi)})\frac{z_k}{\vec{k}^2}\,,\\
&\hat{e}_x:\;\;n_2\simeq 1+b_1 (1-2\Gamma_2\frac{\omega_\phi \xi(\varphi)}{\omega \xi'(\varphi)})\frac{z_k}{\vec{k}^2}\,,\label{dispt2}
\end{align}
with $\varphi=k\cdot x$. However, we argue that the effect would appear as an average over the characteristic formation time of the probe wave. Thus, let's assume
\begin{equation}
\gamma=\frac{1}{T_{f}}\int_0^{T_f}  \frac{\xi(\varphi)}{\xi'(\varphi)} d\tau\,.
\end{equation}
with $T_f$ the formation time of the probe wave. If $\xi$ is a sine or cosine function, as commonly assumed for a long lasting plane wave,  the integral turns out to be zero. But generally speaking, it does not necessarily diminish in other situations. Take for example $\xi=\exp[-\varphi^2/\pi]\cos(\varphi+\theta_\xi)$ as the cases of a few cycle pulse or a sequence of wavepackets. The integral varies $\gamma\sim(-0.1,0.1)$ with different $\theta_\xi$, e.g., $\gamma\sim 0.078$ with $\theta_\xi=-\pi/3$. In the following, when the probe wave/light is mentioned, we mean a few cycle pulse or a wavepacket in a sequence. Therefore, the net effect is not negligible and the indexes of refraction appear as
\begin{align}\label{dispt3}
&\hat{e}_z: n_1\simeq 1+b_2 \mathcal{E}^2-b_1  \gamma \Gamma_2 \mathcal{E}^2 \frac{\omega_\phi }{\omega},\\
&\hat{e}_x: n_2\simeq 1+b_1 \mathcal{E}^2-2 b_1\gamma \Gamma_2 \mathcal{E}^2 \frac{\omega_\phi }{\omega}.\label{dispt4}
\end{align}

The fact that the refraction indexes are different for the two polarization states has long invoked the speculation that it is possible to detect a phenomenon similar to optical birefringence in the QED vacuum, demonstrating the presence of quantum fluctuations \cite{GreinerQED}. A heterodyne detection has been carried out after a plane wave passed through a rotating magnetic field \cite{PVLAS}, an experimental scheme is proposed making use of a high-intensity laser pulse as the pump and a linearly polarized x-ray pulse as the probe \cite{laservb}, and a phase-contrast Fourier imaging technique involving a strong laser pump beam is introduced \cite{Homma2011}.

According to the standard theoretical treatment of the problem \cite{GreinerQED}, the probe light is prepared in a superposition of the two polarization modes as $\sim\sin\theta_0\hat{e}_z \xi(-\omega t+k_1 y)+ \cos\theta_0 \hat{e}_x \xi(-\omega t+ k_2 y)$, with $k_{1,2}=\omega n_{1,2}$ being dependent of time and $\theta_0$ denoting the angle of the polarization vector with respect to $\hat{e}_x$ at $y=0$. Assume the probe light enters the pump field at $y=\mu$.  After moving along the direction of propagation for a distance $L$, the phase difference between the two modes accumulates to be
\begin{equation}\label{phaseacc}
\delta=\int_\mu^{L+\mu} k_1(t_y)-k_2(t_y) dy\,,
\end{equation}
where $t_y=y-\mu$ is the time for the wave plane moving from $l=\mu$ to $l=y$. The ellipticity parameter $\psi=\delta\sin(2\theta_0)/2$ can be experimentally measured. Taking $\theta_0=\pi/4$, then $\psi=\delta/2$.

Substituting Eqs. (\ref{dispt3}, \ref{dispt4}) into Eq. (\ref{phaseacc}), the phase difference can be calculated out. Taking $\mu$ into account is in effect equivalent to introducing a relative phase between the probe light and the pump field, because the value of $\delta$ would be the same if the probe beam enters  at $y=0$ into the pump field of $\mathcal{E}\hat{e}_x\cos(\omega_\phi t+\theta_\phi)$ with $\theta_\phi=\omega_\phi\mu$. The total phase difference reads $\delta=\delta_0/2+\delta'+\delta''$ with
\begin{align}\label{phaseacc2}
&\delta_0=\frac{\alpha\mathcal{E}^2 L}{15 E^2_{cr} \lambda}\,,\\
&\delta'=\frac{\alpha \mathcal{E}^2 \lambda_\phi}{60 \pi E^2_{cr}\lambda}\cos(\omega_\phi L+2\theta_\phi) \sin\omega_\phi L \,,\\
&\delta''=-\frac{\alpha\gamma \mathcal{E}^2}{45 \pi E^2_{cr}}\sin(\omega_\phi L+2\theta_\phi)\sin \omega_\phi L\,,
\end{align}
where $\lambda$ is the wave length of the probe light, $\lambda_\phi$ is that of the pump field and $\delta_0$ is the corresponding phase difference if the pump field is a static field $\mathcal{E}\hat{e}_x$.
The equations show that (1) If $\omega_\phi L\ll 1$, there is
\begin{equation}
\delta \approx \delta_0 \cos^2 \theta_\phi.
\end{equation}
This is what to be expected, since the pump field varies so slowly during the propagation of the probe wave that it shall be seen as a constant field with the amplitude $\mathcal{E} \cos\theta_\phi$ at the incidence. (2) For a notable value of $\omega_\phi$, $\delta'$ and $\delta''$ oscillate as $L$ varies. The presumption that the probe beam is shot at the antinode of the pump field implies that $\lambda_\phi\gg \lambda$ as discussed before. Taking $\gamma\sim 0.1$, the amplitude of $\delta'$ can be much larger than that of $\delta''$.
For $\theta_\phi=0$, the maximum value of the ellipticity parameter can be obtained as
\begin{equation}
\psi_{\max}=\frac{\alpha \mathcal{E}^2}{2 E^2_{cr}\lambda}(\frac{L}{30}+\frac{\lambda_\phi}{120\pi})\,.
\end{equation}
(3) If $\theta_\mu$ is not controlled and varies randomly among different shots, to avoid its influence on the result it is favored to set the propagation length to be $L=n\pi/\omega_\phi$ with $n=1,2,\cdots$, and thus $\delta'$ as well as $\delta''$ is suppressed. Otherwise the measurement data would randomly oscillate around the value $\delta_0/2$ from shot to shot.



It is recognized that $\delta'$ comes from the terms $\sim b_{1,2}z_k=c_{1,2}z_k\cos^2\omega_\phi t$ in the relations (\ref{disp_eq2}, \ref{disp_eq5}), which can be seen as direct generalizations of the terms $\sim c_{1,2}z_k$ in the conventional light cone conditions (\ref{lightcone1}, \ref{lightcone2}) through replacing the constant field $\mathcal{E}$ in the latter case by the field value $\mathcal{E}\cos\omega_\phi t$ at time $t$. However, $\delta''$ originates from the additional corrections $~\sim b_{1,2}\Gamma_1\Gamma_2\omega_\phi/\omega$ in the relations (\ref{disp_eq2}, \ref{disp_eq5}) which have no correspondence in (\ref{lightcone1}, \ref{lightcone2}).
It is interesting to note that the amplitude of $\delta''$ does not depend on $\omega$ or $\omega_\phi$, although the corrections $~\sim b_{1,2}\Gamma_1\Gamma_2\omega_\phi/\omega$ is explicitly proportional to the ratio $\omega_\phi/\omega$ at any given time. Instead, it contains the factor $\gamma$ which illustrates that the effect is an average over the formation time of the probe light, echoing with the explanation at the end of the previous section.


The amplitude of $\delta''$ differs from $\delta_0$ by an order of $\gamma\lambda/L$. In experiments like those described in \cite{PVLAS}, the effective propagation length can be increased to be over $10^4$m, so for an optical light the ratio $\gamma\lambda/L$ can be as small as $10^{-11}$ and $\delta''$ is negligible in measurements. However, in experiments involving intense lasers like those proposed in \cite{Homma2011, Benking, laservb},  the effective length $L$ is in the order of the waist size of the pump laser beam. Experimentally the laser beam can be focused to the spot size within a few times of its wavelength.  As a conservative estimation, suppose $L/\lambda_\phi\sim 10$, $\lambda_\phi/\lambda\sim100$ to insure the antinode assumption is applicable and $\gamma\sim0.1$, the ratio $\gamma\lambda/L\sim 10^{-4}$ is largely enhanced. The constraint with regard to the antinode assumption may be loosened, by noticing that the phase difference of the same probe light shooting at the magnetic antinode of the same pump field turns out to be
\begin{equation}
\delta=-\frac{\alpha\gamma \mathcal{E}^2}{45 \pi E^2_{cr}}\sin(\omega_\phi L+2\theta_\phi)\sin \omega_\phi L\,,
\end{equation}
which is exactly the same as $\delta''$. This supports the speculation that a phase difference like $\delta''$ shows up for arbitrary choice of the incidence point or for arbitrary size of the crosssection of the probe light. If this is indeed the case, then the requitement $\lambda_\phi\gg\lambda$ can be lifted, a correction like $\delta''$ generally exists, and the ratio $\gamma\lambda/L$ can be further raised.

Envisaging the advanced strong laser with intensity $\sim 10^{25}$W/cm$^2$ as the pump field, the absolute magnitude of $\delta''$ could be $\sim5\times10^{-10}$, approaching the sensitivity of high-precision polarimetry technique available nowadays \cite{sensitivity}. Besides, an important object of studying vacuum birefringence is to search for signals from new physics, which are likely to be much weaker than the signal of $\sim \delta_0$. If it is in the same order of $\delta''$ or weaker, the knowledge about $\sim\delta''$ is required to extract useful information from the measured data.

It is worthy to emphasize that the above derivations are performed in the regime of the lowest-order Heisenberg-Euler Lagrangian.  However, the temporal and/or spatial dependency of the fields can invoke extra corrections involving derivatives of the field strength tensors to the lagrangian. The extra terms are given by \cite{Markuland_photonacceleration, Mamaev_corr}
\begin{equation}
\delta\mathcal{L}'=\frac{2\epsilon_0\alpha c^2}{15\omega_c^2}(\partial_\mu F^{\mu\nu}\partial_\lambda F^\lambda_{\;\;\nu}-F_{\mu\nu}\partial_\lambda\partial^\lambda F^{\mu\nu})\,.
\end{equation}
It is found that this would lead to an extra correction of the light cone condition proportional to $\alpha^2(\omega/\omega_c)^2(|\vec{E}|/E_{cr})^2)$ \cite{Rozanov1998, Markuland_photonacceleration}.
In the measurement of the ellipticity parameter, this results in a correction by the factor $\alpha(\omega/\omega_c)^2$ smaller than $\sim\delta_0$, and by the factor $\alpha L/(\gamma\lambda)(\omega/\omega_c)^2$ different from $\sim\delta''$. Therefore, the comparison between this correction and $\sim\delta''$ should be assessed case by case.
For the situation where the effective propagation length is as long as $\sim 10^4$m \cite{PVLAS}, if the probe light is optical $\omega\sim1$eV, this correction is smaller than $\sim\delta''$ by an order of $10^{-3}$, and if the probe light is hard x ray lasers under development $\omega\sim 10$keV \cite{XFEL}, the correction is larger than $\delta''$ by an order of $10^5$. But even under such high frequency condition, for the aforementioned strong laser pump field setups with short effective propagation length, it could still be several orders smaller than $\sim\delta''$.

\subsection{Summary and outlook}
Vacuum polarization by a pump field composed of two counterpropagating laser beams is studied. The light cone condition of a probe wave propagating in the excited vacuum, especially on the electric/magnetic antinode plane of the pump field, is obtained based on the lowest-order Heisenberg-Euler lagrangian. The differences to the commonly referred light cone conditions in a constant pump field are notable. Besides a direct generalization of the conventional relations, particularly an additional correction exists. However, the analysis of birefringence measurement shows that for most cases this correction contributes a small fraction to the ellipticity parameter. Nonetheless, with the presence of intense laser fields this additional correction could lead to detectable signals in a highly sensitive polarimeter which might mix with or even cover the expected heralding signals of new physics.

The analytical derivation relies on the particular field configuration. It is postulated that on the path of the probe wave the pump field is a time dependent electric/magnetic field while the amplitude of the magnetic/electric field vanishes. Moreover, real beam effects, such as beam focusing and Gouy phase shift, are not taken into account.  A universal light cone condition is not acquired yet. In view of the inexhaustible possibilities of different situations,  it is speculated that numerical calculations like those performed in \cite{Benking} would play a crucial role in exploring vacuum polarization problems in diverse sorts of field configurations.

We are indebted to Professor C. S. Lam and Professor C. M{\"u}ller for useful discussions and valuable advices and acknowledge financial support from the National Natural Science Foundation of China under Grant No. 11204370.

\end{document}